\documentclass[prl,twocolumn,aps,superscriptaddress,showpacs,preprintnumbers,nofootinbib]{revtex4}

\usepackage{varioref,exscale,latexsym,amsmath,amssymb}
\usepackage{graphicx}
\usepackage{bm}
\usepackage{slashed}
\usepackage[raggedright]{titlesec}

\begin{document}
\title{The Higgs Mass and the Emergence of New Physics}

\author{Ufuk Aydemir}\email{uaydemir@vt.edu}
\affiliation{Department of Physics, Virginia Tech, Blacksburg, VA 24061, USA}

\author{Djordje Minic}\email{dminic@vt.edu}
\affiliation{Department of Physics, Virginia Tech, Blacksburg, VA 24061, USA}

\author{Tatsu Takeuchi}\email{takeuchi@vt.edu}
\affiliation{Department of Physics, Virginia Tech, Blacksburg, VA 24061, USA}
\affiliation{Kavli Institute for the Physics and Mathematics of the Universe (WPI),
The University of Tokyo, Kashiwa-shi, Chiba-ken 277-8583, Japan}

\date{\today}

\begin{abstract}
We investigate the physical implications of formulating the electroweak (EW) part of the Standard Model (SM)
in terms of a superconnection involving the supergroup $SU(2/1)$.
In particular, we relate the observed Higgs mass to new physics at around $4$~TeV.
The ultraviolet incompleteness
of the superconnection approach points to its emergent nature.
The new physics beyond the SM is associated with the emergent supergroup
$SU(2/2)$, which is natural from the point of view of the Pati-Salam model.
Given that the Pati-Salam group is robust in certain constructions of string vacua,
these results suggest a deeper connection between low energy ($4$~TeV) and high energy (Planck scale) physics
via the violation of decoupling in the Higgs sector.
\end{abstract}

\pacs{11.10.Hi,11.10.Nx,11.30.Pb,12.90.+b}

\preprint{IPMU13-0052}

\maketitle
\section{\normalsize{Introduction and Overview}}
The Standard Model (SM) of particle physics is a phenomenally successful theory whose
last building block has recently been detected \cite{atlas:2012gk,cms:2012gu}.
In light of the apparent discovery of the Higgs boson, 
we address the connection between its mass and the structure of the
electroweak (EW) sector of the SM, and argue that it points to
some very exciting new physics at a rather low energy scale of $4$ TeV.

A long time ago, Ne'eman \cite{Neeman:1979wp} and Fairlie \cite{Fairlie:1979at} independently discovered the relevance of a unique $SU(2/1)$ supergroup structure to SM physics. In this formalism, the even (bosonic) part of the $SU(2/1)$ algebra defines the $SU(2)\times U(1)$ gauge sectors of the SM, while the Higgs sector is identified as the odd (fermionic) part of the algebra. Although the model gives the correct quantum numbers of the SM, and it represents a more unified-hence more aesthetic-version of the SM, it suffers from the violation of the spin-statistic theorem, a common problem seen in the models using supergroups.\footnote{For example, there are anticommuting Lorentz scalars (the Higgs fields) which represent ghost-like degrees of freedom in the model.}

In this work we adopt the superconnection approach of Ne'eman and Sternberg \cite{Ne'eman:1990nr} who observed
that the $SU_L(2)\times U_Y(1)$ gauge and Higgs bosons of the SM
could be embedded into a unique $SU(2/1)$ superconnection, and the quarks and leptons into
$SU(2/1)$ representations \cite{Ne'eman:2005cy, Hwang:1995wk}. $SU(2/1)$ in this formalism is not imposed as a symmetry; it is rather only the structure group of the superconnection. Therefore, the $SU(2/1)$ structure can be interpreted as an emergent geometric pattern that involves the EW part of the SM, which avoids the problems with the ghosts.

The formalism fixes the ratio of the $SU_L(2)\times U_Y(1)$ gauge couplings,
and thus the value of $\sin^2\theta_W$,
and the quartic coupling of the Higgs.
The value of $\sin^2\theta_W$ selects the scale $\Lambda\sim 4\,\mathrm{TeV}$
at which the superconnection relations can be imposed\footnote{This scale is updated from the 5~TeV in Ref.~\cite{Ne'eman:2005cy}
using more recent determinations of the gauge couplings.
The difference does not play a noticeable role in the prediction of the Higgs mass.}, and renormalization group (RG) running
leads to a prediction of the Higgs mass.
However, 
the claim of Refs.~\cite{Ne'eman:2005cy,Hwang:1995wk} that
the predicted Higgs mass is around $130$~GeV turns out to be incorrect.

In this Letter, we point out that the $SU(2/1)$ superconnection approach
predicts the mass of the Higgs to be $170$~GeV,
which obviously disagrees with observation.
Given the well-known issue with the ultraviolet incompleteness of the $SU(2/1)$ approach \cite{Ne'eman:2005cy},
which implies the emergent nature of this description,
we should have no qualms in introducing
new physics to fix the Higgs mass.

Here, we note a connection with the Spectral SM of Connes and collaborators \cite{connes:1994,connes:2007}
in which spacetime is extended to a product of a continuous four dimensional manifold
by a finite discrete space with non-commutative geometry.
The SM particle content and gauge structure are described by a unique geometry, where
the Higgs appears as the connection in the extra discrete dimension \cite{Coquereaux:1990ev}.
Curiously, the original Higgs mass prediction of the Spectral SM was also 170~GeV \cite{Chamseddine:2010ud}, despite the fact that the boundary conditions imposed on the RG equations were quite different:
in the Spectral SM, the usual $SO(10)$ relations among the gauge couplings are imposed at the GUT scale.
In a recent paper \cite{spectral} Chamseddine and Connes isolate a unique scalar
degree of freedom that is responsible for the neutrino Majorana mass in their approach, which, when correctly coupled to the Higgs field, can reduce the mass of the Higgs boson to the observed value, $125\sim 126$~GeV.\footnote{Given the similarities between the outcomes of the Spectral Model of Connes and Chamseddine \cite{spectral} and the superconnection formalism, there may be a relation between these models.}

We argue that
a similar `fix' works for the superconnection formalism:
one needs to introduce extra scalar degrees of freedom which modify the RG equations.
We further point out that this can be accomplished by the embedding of $SU(2/1)$ into $SU(2/2)$,
and thus, in effect, a left-right (LR) symmetric extension of the EW sector \cite{leftright}, which is also natural from the point of view of the Pati-Salam model \cite{Pati:1974yy}. The $SU(2/2)$ formalism, as in the $SU(2/1)$ case, selects the scale $\Lambda\sim 4\,\mathrm{TeV}$ via the value of $\sin^2\theta_W$. Therefore, $4\,\mathrm{TeV}$ in this formalism is the \emph{prediction} for the energy scale of \emph{new physics}, which is the LR symmetric model in this case.



We also note the peculiarity of the Higgs sector,
which due to the relation between the coupling and the mass, 
violates decoupling \cite{nondec}. When interpreted from either the emergent superconnection
or the non-commutative geometry viewpoint,
this violation of decoupling offers an exciting connection between the SM and short distance physics,
such as string theory, via the non-decoupling of the 4~TeV and the Planck scales.

In particular, the embedding of $SU(2/1)$ into $SU(2/2)$ would be interesting from the point of view of
string vacua, where it has been observed that the Pati-Salam group appears rather ubiquitously
in a large number of vacua \cite{dienes}.
Though we lack a fundamental understanding of this phenomenon,
it is quite intriguing in our context as it would point to a
new relationship between low energy (SM-like) and high energy physics (quantum-gravity-like) which
is not seen in the standard effective field theory approach to particle physics.

\section{\normalsize{ The ${\textbf{SU(2/1})}$ formalism and the Higgs mass}}
Here we summarize the superconnection approach to the SM based on the supergroup $SU(2/1)$ \cite{Ne'eman:2005cy,Hwang:1995wk}. Obviously, this supergroup has as its bosonic subgroup the EW gauge group $SU(2)_L\times U(1)_Y$.
What is highly non-trivial is that
the embedding of $SU(2)_L\times U(1)_Y$ into $SU(2/1)$ also gives the correct quantum numbers for all the physical
degrees of freedom.
Furthermore, the Higgs sector comes out naturally as a counterpart of the gauge sector.
These have natural analogs in the Spectral SM as well \cite{connes:1994,connes:2007,spectral},
as already emphasized in the conclusion to the review Ref.~\cite{Ne'eman:2005cy}.
We concentrate on the superconnection formalism which should be understood as an emergent framework,
because of the fundamental ultraviolet incompleteness of gauged supergroup theories.

We start by defining the supercurvature as
$
\mathcal{F}=\textbf{d}\,\mathcal{J}+\mathcal{J}\cdot\mathcal{J}
$
where $\mathcal{J}$ is the superconnection, which is of the form
\begin{equation}
\mathcal{J}\;=\;
\left[
              \begin{array}{cc}
                M & \phi \\
                \overline{\phi}& N \\
              \end{array}
\right].
\end{equation}
Since we would like to embed $SU_L(2)\times U_Y(1)$ and the Higgs into $SU(2/1)$,
$M$ and $N$ are respectively $2\times 2$ and $1\times 1$ 
g-even submatrices valued over one-forms,
while $\phi$ and $\overline{\phi}$ are respectively $2\times 1$ and $1\times 2$ 
g-odd submatrices valued over zero-forms.
The superconnection $\mathcal{J}$ is written as $\mathcal{J}=i\lambda_s^a J^a$, $a=1,2,\cdots,8$. The generators $\lambda_s^a$ are matrices with supertrace zero.
Therefore, they are the usual $SU(3)$ $\lambda$-matrices except for $\lambda_s^8$ which is
\begin{equation}
\lambda^8_s\;=\;
\frac{1}{\sqrt{3}}\left[
              \begin{array}{ccc}
                -1 & 0 & 0 \\
                0 & -1 & 0 \\
                0 & 0 & -2 \\
              \end{array}
            \right].
\end{equation}
To obtain the superconnection we need to make the identifications $J^i=W^i$ $(i=1,2,3)$ and $J^8=B$, where $W^i$ and $B$ are one-form fields corresponding to the $SU_L(2)$ and $U_Y(1)$ gauge bosons.
The zero-form fields are identified as $J^4\mp iJ^5=\sqrt{2}\phi^\pm$, $J^6 - iJ^7=\sqrt{2}\phi^0$, and $J^6 + iJ^7=\sqrt{2}\phi^{0\ast}$.\footnote{Note that $^\ast$, which we will use to denote the Hodge product later in the paper, here denotes taking complex conjugate of a field.
}
Then, the superconnection is
\begin{equation}
\mathcal{J}\;=\;
i\left[
  \begin{array}{cc}
   \mathcal{W}-\frac{1}{\sqrt{3}}B\cdot \textbf{I} & \sqrt{2}\Phi \\
    \sqrt{2}\Phi^\dagger & -\frac{2}{\sqrt{3}}B \\
  \end{array}
\right].
\end{equation}
Here, $\mathcal{W}=W^i\,\tau^i$ (where $\tau^i$ are the Pauli matrices) and $\textbf{I}$ is a $2\times2$ unit matrix, and
$
\Phi =\left[                                                                                                                            \phi^+ \; \phi^0                                                                                                                                                                                                                                \right]^\mathrm{T}.
$
To obtain the supercurvature $\mathcal{F}$, we recall the rule for supermatrix multiplication \cite{Ne'eman:1990nr, Hwang:1995wk}
\begin{eqnarray}\label{rule}
\lefteqn{
\left[
  \begin{array}{cc}
    A & C \\
    D & B \\
  \end{array}
\right]\cdot
\left[
  \begin{array}{cc}
    A' & C' \\
    D' & B' \\
  \end{array}
\right]}
\cr
& \!\!\!=\!\!\! &
\left[\!
  \begin{array}{cc}
    A\wedge A'+(-1)^{|D'|} C\wedge D' &\,  A\wedge C'+(-1)^{|B'|}C\wedge B' \\
    (-1)^{|A'|} D\wedge A'+B\wedge D' &\, (-1)^{|C'|} D\wedge C'+B\wedge B' \\
  \end{array}
\!\right]
\cr
& &
\end{eqnarray}
%
where $|A|$ denotes the $Z_2$ grading of the differential form $A$.
Then, the supercurvature (after introducing the dimensionless coupling $g$, $\mathcal{J}\rightarrow g\mathcal{J}$) reads as
\begin{equation}
\mathcal{F}\;=\;i g
\left[
              \begin{array}{cc}
                F_W-\frac{1}{\sqrt{3}}F_B\cdot\textbf{I}+2ig\Phi\Phi^{\dagger} & \sqrt{2}D\Phi \\
                \sqrt{2}(D\Phi)^{\dagger} & -\frac{2}{\sqrt{3}}F_B+2ig\Phi^{\dagger}\Phi \\
              \end{array}
\right]
\end{equation}
where
$
D\Phi=d\Phi+ig\mathcal{W}\Phi+ig\frac{1}{\sqrt{3}}B\Phi$, $F_B=dB$ and
$F_W=\left(F_W\right)^k\tau^k=\left[d W^k+ig\,\epsilon^{ijk}W^i\wedge W^j\right]\tau^k$.
The Action reads as follows
\begin{eqnarray}
\mathcal{S}
& = &
\int\frac{-1}{4g^2}\mbox{Tr}\left[\mathcal{F}\cdot\mathcal{F}^\star\right]
\cr
& = &
\int\bigg(\frac{1}{2}
\left[
-\left(F_W\right)^i\wedge\left(F_W^\ast\right)^i-
F_B\wedge F_B^\ast
\right]
\cr
& &
+\left(D\Phi\right)^{\dagger}\wedge \left(D\Phi\right)^\ast-\lambda\left(\Phi^{\dagger}\Phi\right)\wedge \left(\Phi^{\dagger}\Phi\right)^\ast\bigg),
\end{eqnarray}
where the $\star$ on $\mathcal{F}^\star$ denotes taking the Hermitian conjugate of the supermatrices and the Hodge dual (denoted as $\ast$) of the differential forms, and
$\lambda\equiv 2g^2$. Note that we need to break $SU(2/1)$ explicitly in order to introduce the Higgs mass.
In 4 dimensions  we have the following explicit form of the Lagrangian
(given the metric $g_{\mu\nu}=\mathrm{diag}(1,-1,-1,-1)$):
\begin{eqnarray}
\mathcal{L}
& = & -\frac{1}{4}F_{W\,\mu\nu}^i\,F^{i\,\mu\nu}_W-\frac{1}{4}F_{B\,\mu\nu}F_B^{\mu\nu}
\cr
& &
+\left(D_{\mu}\Phi\right)^{\dagger}\left(D^{\mu}\Phi\right)
-\lambda\left(\Phi^{\dagger}\Phi\right)^2.
\end{eqnarray}
Note that the explicit forms of the curvature strengths and the covariant derivatives have the standard forms:
$
F_{W\mu\nu}^i=\partial_{\mu}W^{i}_{\nu}-\partial_{\nu}W^{i}_{\mu}+2ig\epsilon^{jki} W^{j}_{\mu}W^{k}_{\nu}$,
$F_{B\mu\nu}=\partial_{\mu}B_{\nu}-\partial_{\nu}B_{\mu}$ and
$D_{\mu}\Phi=\partial_{\mu}\Phi+ig\left(\boldsymbol{\tau}\cdot\mathbf{W}_{\mu}\right)\Phi+ig'B_{\mu}\Phi$, with
$g'/g=1/\sqrt{3}$.
To switch to the common SM convention we rescale $g$ and $g'$ as $g,g\rightarrow g/2, g'/2$ (which is the missing part in \cite{Hwang:1995wk}) which also changes our constraint at the symmetry breaking energy to $\lambda=g^2/2$.\footnote{If we do not make these rescalings at this point then we need to make appropriate ones in Eq. (\ref{RGLang}).} Now we address the prediction for the Higgs mass.
In what follows we use the relation
$M_H^2=8M_W^2(\lambda/g^2)$ and the RG equations for $\lambda$ and top Yukawa coupling $g_t$ which are
\begin{eqnarray}\label{RGLang}
\mu\frac{d h_t}{d\mu}
&=&\frac{h_t}{(4\pi)^2}\left(\frac{9}{2}h_t^2-\left(\frac{17}{12}g'^2+\frac{9}{4}g^2+8g_s^2\right)\right),\cr
\mu\frac{d \lambda}{d\mu}
&=&\frac{1}{(4\pi)^2}\bigg(\left(12 h_t^2-\left(3 g'^2+9 g^2\right)\right)\lambda-6 h_t^4\cr
&+&24\lambda^2+\frac{3}{8}\left(g'^4+2g'^2 g^2+3g^4\right)\bigg),
\end{eqnarray}
where $g'$, $g$, and $g_s$ are the $U(1)_Y$, $SU(2)_L$ and $SU(3)_c$ coupling constants,  respectively,
$h_t=\sqrt{2}M_t/v$, and $M_t=173.4 \,\,\mbox{GeV}$ is the mass of the top quark.
We will follow Ref.~\cite{Neeman:1979wp} to find the boundary condition on $\lambda$.
To find the scale of emergence of $SU(2/1)$ ($\Lambda_s$), we find the scale where the group theoretical value for $\theta_W$, $g=\sqrt{3}g'$ ($\sin^2{\theta_W}=0.25$), holds. We use
\begin{equation}
\frac{1}{[g_i(\Lambda_s)]^2}\;=\;\frac{1}{[g_i(\Lambda_0)]^2}-2 b_i\ln\frac{\Lambda_s}{\Lambda_0}\; \quad (i\;=\;1,2,3)
\end{equation}
where the respective constants $b_i$ read as:
\begin{eqnarray}
b_1 & = &  \frac{1}{16\pi^2}\left(\frac{20\,n_f}{9}+\frac{n_H}{6}\right)\;,\cr
b_2 & = & -\frac{1}{16\pi^2}\left(-\frac{4\,n_f}{3}-\frac{n_H}{6}+\frac{22}{3}\right)\;,\cr
b_3 & = & -\frac{1}{16\pi^2}\left(-\frac{4\,n_f}{3}+11\right)\;.
\end{eqnarray}
Setting the number of fermion families to $n_f=3$, and the number of Higgs doublets to $n_H=1$,
we find $\Lambda_s\simeq4$ TeV (note that $g_1=g'$, $g_2=g$, $g_3=g_s$).
Using Eq.~(\ref{RGLang}) with the boundary conditions $\lambda=g_2^2/2$ at 4 TeV and
$h_t=\sqrt{2}M_t/v$ at $M_Z$, we find that $\lambda(M_Z)\simeq0.24$ and thus $M_H\simeq 170$ GeV.  The numerical values ($\overline{\mbox{MS}}$) we use in this calculation \cite{Beringer:1900zz} are $\alpha_1^{-1}(M_Z)=98.36$, $\alpha_2^{-1}(M_Z)=29.58$, $\alpha_3^{-1}(M_Z)=8.45$, where $\alpha_i^{-1}=4\pi/g_i^2$.


\section{ \normalsize{The ${\textbf{SU(2/2)}}$ embedding}}
Given the incorrect mass of the Higgs and the fact that the superconnection approach suffers from ultraviolet incompleteness, and
thus it has to be considered only as an emergent description, we now introduce new emergent physics to correct the Higgs mass.
In this section, we use $SU(2/2)$ instead of $SU(2/1)$ to do the embedding.
(From the Spectral SM viewpoint $SU(2/2)$ would correspond to a symmetric non-commutative geometry.)
In this case, the embedded gauge group is $SU(2)_L\times SU(2)_R\times U(1)_{B-L}$.
We follow the same route as in the previous section and find the energy scale of the new physics predicted by this structure.
We also make the simplifying assumption that this energy scale is also the energy scale
at which $SU(2)_L\times SU(2)_R\times U(1)_{B-L}$ breaks to the SM.
First, we find the superconnection we need.
Given the generators of $SU(2/2)$,
$\mathcal{J}$ can be expressed as $\mathcal{J}=i\lambda^a_s\,J^a$, $a=1,2,\cdots,15$.
We make the following identifications:
$J^{1,2,3}=W_L^{1,2,3}$, $J^{13,14,8}=W_R^{1,2,3}$,
$J^4-iJ^5=\sqrt{2}\phi_1^0$, $J^4+iJ^5=\sqrt{2}\phi_1^{0\ast}$,
$J^6-iJ^7=\sqrt{2}\phi_2^-$, $J^6+iJ^7=\sqrt{2}\phi_2^+$, $J^9-iJ^{10}=\sqrt{2}\phi_1^+$,
$J^9+iJ^{10}=\sqrt{2}\phi_1^-$,
$J^{11}-iJ^{12}=\sqrt{2}\phi_2^0$ and
$J^{11}+iJ^{12}=\sqrt{2}\phi_2^{0\ast}$.
Here $W_L^i$ and $W_R^i$ are 1-forms and the others are 0-form fields
corresponding to the left- and right-handed gauge bosons and the bidoublet Higgs field.
As a result, we obtain the superconnection, a $4\times 4$ supermatrix, in the following form
\begin{eqnarray}\label{superconnection2}
\mathcal{J}=i\left[
              \begin{array}{cc}
                W_L-\frac{1}{\sqrt{2}}W_{BL}\cdot \textbf{I} & \sqrt{2}\Phi \\
                \sqrt{2}\Phi^{\dagger} & W_R-\frac{1}{\sqrt{2}}W_{BL}\cdot \textbf{I} \\
              \end{array}
            \right]
\end{eqnarray}
where
\begin{equation}
W_L = W_L^i\tau^i\;,\quad
W_R = W_R^i\tau^i\;,\quad
\Phi = \left[
\begin{array}{cc}
\phi_1^0 & \phi_1^+ \\
\phi_2^- & \phi^0_2 \\
\end{array}
\right]\;.
\end{equation}
This leads to the following expression for $\mathcal{F}$ (after rescaling $\mathcal{J}$ as $g\mathcal{J}$)
\begin{equation}
\mathcal{F}
= ig
\left[
               \begin{array}{cc}
                 F_L -\frac{1}{\sqrt{2}}\widetilde{F}_{BL}+2ig\Phi\Phi^{\dagger}& \sqrt{2}D\Phi \\
                 \sqrt{2}(D\Phi)^{\dagger} & F_R-\frac{1}{\sqrt{2}}\widetilde{F}_{BL}+2ig\Phi^{\dagger}\Phi \\
               \end{array}
\right]
\end{equation}
where $W_{LR}\,=\,W_{LR}^i\,\tau^i$, $\,\widetilde{F}_{BL}=F_{BL}\cdot \textbf{I}\,=\,d W_{BL}\cdot \textbf{I}$, \\ $F_{L,R}=(F_{L,R})^a\,\tau^a=(d W_{L,R}^i+ig(W_{L,R}\wedge W_{L,R})^i)\,\tau^i$, and
$D\Phi=d\Phi+igW_L\Phi-ig W_R\Phi$.
The corresponding action $\mathcal{S}=\int\frac{-1}{4g^2}\mbox{Tr}\left[\mathcal{F}\cdot\mathcal{F}^\star\right]$ now reads as
\begin{eqnarray}\label{lagrangian2}
\lefteqn{\mathcal{S}
\;=\;\int-\bigg(\frac{1}{2}\left(F_L\right)^i\wedge\left(F_L^\ast\right)^i+\frac{1}{2}\left(F_R\right)^i\wedge\left(F_R^\ast\right)^i} \cr
&& +\frac{1}{2}F_{BL}\wedge F_{BL}^\ast-\mbox{Tr}\left[\left(D\Phi\right)^{\dagger}\wedge \left(D\Phi\right)^\ast\right]\cr
&& +\widetilde{\lambda}\left\{\mbox{Tr}\left[\left(\Phi^{\dagger}\Phi\right)\wedge \left(\Phi^{\dagger}\Phi\right)^\ast\right]+\mbox{Tr}\left[\left(\Phi\Phi^{\dagger}\right)\wedge \left(\Phi\Phi^{\dagger}\right)^\ast\right]\right\}\bigg)
\cr
&&
\end{eqnarray}
where $\lambda\equiv g^2$. In 4 dimensions, with the metric $g_{\mu\nu}=\mathrm{diag}(1,-1,-1,-1)$, the Lagrangian (again with the rescaling $g\rightarrow g/2$) becomes
\begin{eqnarray}
\lefteqn{\mathcal{L}\;=\;-\frac{1}{4}F_{L\,\mu\nu}^i\,F^{i\,\mu\nu}_L-\frac{1}{4}F_{R\,\mu\nu}^i\,F^{i\,\mu\nu}_R-\frac{1}{4}F_{BL\,\mu\nu}\,F^{\mu\nu}_{BL}}\cr
&&+\mbox{Tr}\left[\left(D_{\mu}\Phi\right)^{\dagger} \left(D^{\mu}\Phi\right)\right]
-\widetilde{\lambda}\mbox{Tr}\left[\left(\Phi^{\dagger}\Phi\right)^2+\left(\Phi\Phi^{\dagger}\right)^2\right].\nonumber\\
\end{eqnarray}
where now $\widetilde{\lambda}= g^2/4$.
To relate $\widetilde{\lambda}$ to the SM $\lambda$, we look at the potential term at the symmetry breaking scale where
$\Phi$ acquires vacuum expectation values (VEVs)
$\langle\Phi\rangle=\left(
       \begin{array}{cc}
         \kappa & 0 \\
         0 & \kappa' \\
       \end{array}
     \right)$,
so that $V(\langle\Phi\rangle)=2\widetilde{\lambda}\left(|\kappa|^4+|\kappa'|^4\right)$.
We equate this to $V_{SM}=\lambda\,v^4/4$, where
$v/\sqrt{2}=\left(|\kappa|^2+|\kappa'|^2\right)^{1/2}$ \cite{Langacker:2010zza}. Assuming $|\kappa|\gg |\kappa'|$,\footnote{Either $\kappa$ or $\kappa'$ must be very small or vanishing as required by the suppression of the flavor changing neutral-currents (FCNC) \cite{Deshpande:1990ip}.} we find
$\lambda\cong 2\widetilde{\lambda}$, and the constraint becomes $\lambda=g^2/2$,
which is the same as that for the $SU(2/1)$ case.
Similarly, the prediction for $\sin^2{\theta_W}$ can be shown to be the same as
in the $SU(2/1)$ case \cite{ournext}.

The $SU(2/2)$ structure has to be broken explicitly in order to introduce the Higgs mass, which is similar to the $SU(2/1)$ case. Additionally, we need to introduce extra scalars in the triplet representation of $SU(2)_{L,R}$ which are necessary in LR symmetric models in order to break the $SU(2)_R\times U(1)_{B-L}$ to $U(1)_Y$ by appropriate VEVs.\footnote{This can be accomplished by a doublet as well. The advantage of the triplet representation is that it can yield a Majorana mass term for the right-handed neutrino.} These triplets may be remnants of a larger geometrical structure, e.g. $SU(N/M)$.

\section{\normalsize{The observed Higgs boson mass from ${\textbf{SU(2/2)}}$}}
\begin{figure}[t]
\leftline{
\includegraphics[width=0.47\textwidth]{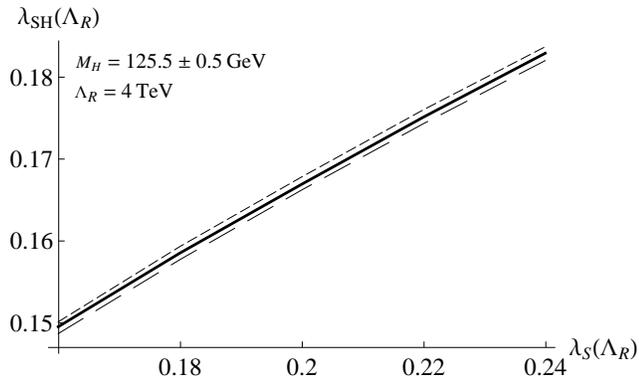}
}
\caption{A patch of the parameter space which gives the observed Higgs mass.}
\label{lmvalues}
\end{figure}

Let us now discuss how the observed Higgs mass comes about.
We have seen that both $SU(2/1)$ and $SU(2/2)$ embeddings
predict the scale of new physics as $\sim 4$~TeV,
provided in the latter case that the $SU(2/2)$ emerges at the same scale as where the $SU(2)_L\times SU(2)_R\times U(1)_{B-L}$ breaks down to $SU(2)_L\times U(1)_{Y}$.
Moreover, they require the same boundary condition at 4~TeV.
This makes the $SU(2/2)$ embedding more appealing since there are a variety of terms that can bring the Higgs mass down to its measured value.
In this section, we will investigate the simplest option as an example. We will assume that only a scalar singlet survives dominantly at low energies ($\sim M_Z$) which is responsible for the mass of the right-handed neutrino and which comes out naturally in the Spectral SM \cite{spectral}.
The model in which the SM is extended with a scalar has been worked out before in detail in the contexts of vacuum stability of the SM \cite{vacuumstability} and dark matter \cite{darkmatter}. We will explore the parameter space of this model in the framework of $SU(2/2)$.
The RG equations can be written as
\begin{eqnarray}
&&\mu\frac{d h_t}{d\mu}=\frac{h_t}{(4\pi)^2}\left(\frac{9}{2}h_t^2+ h_{\nu}^2-\left(\frac{17}{12}g'^2+\frac{9}{4}g^2+8g_s^2\right)\right),\cr
&&\mu\frac{d h_{\nu}}{d\mu}=\frac{h_{\nu}}{(4\pi)^2}\left(3h_t^2+\frac{5}{2}h_{\nu}^2-\left(\frac{3}{4}g'^2+\frac{9}{4}g^2\right)\right),\cr
&&\mu\frac{d \lambda}{d\mu}=\frac{1}{(4\pi)^2}\bigg(\left(12 h_t^2+4 h_{\nu}^2-\left(3 g'^2+9 g^2\right)\right)\lambda-2h_{\nu}^4\cr
&&-6 h_t^4+2\left(12\lambda^2 +\lambda_{HS}^2+\frac{3}{16}\left(g'^4+2g'^2 g^2+3g^4\right)\right)\bigg),\cr
&&\mu\frac{d \lambda_{HS}}{d\mu}=\frac{\lambda_{HS}}{(4\pi)^2}\bigg(6h_t^2+2h_{\nu}^2-\frac{3}{2}g'^2-\frac{9}{2}g^2\cr
&&\,\,\,\,\,\,\,\,\,\,\,\,\,\,\,\,+2\left(6\lambda+3\lambda_S+4\lambda_{HS}\right)\bigg),\cr
&&\mu\frac{d \lambda_{S}}{d\mu}=\frac{1}{(4\pi)^2}\left(8\lambda_{HS}^2+18\lambda_{S}^2\right),
\end{eqnarray}
where $h_t$ and $h_{\nu}$ are the top-quark and right-handed neutrino Yukawa couplings, $\lambda$ and $\lambda_S$ are the Higgs and the singlet quartic couplings, and $\lambda_{HS}$ is the Higgs-singlet coupling. The boundary conditions we use are $h_t(M_Z)=0.997$, obtained from $h_t(M_Z)=\sqrt{2}M_t/v$, and
$\lambda(\Lambda_R)=g^2(\Lambda_R)/2$,
where the latter is fixed by the $SU(2/2)$ construction.
We also assume $h_{\nu}\sim 10^{-6}$, which is necessary to generate the correct light neutrino mass from the TeV scale seesaw, if the Dirac mass $M_D\approx M_e$. There are still two more boundary conditions, corresponding to ones on  $\lambda_S(\Lambda_R)$ and $\lambda_{HS}(\Lambda_R)$, which are \emph{not} fixed by $SU(2/2)$. The mass of the Higgs can be determined by using \cite{spectral}
\begin{eqnarray}
M_H^2
& = & \lambda v^2+\lambda_S v_R^2-\sqrt{\left(\lambda v^2-\lambda_S v_R^2\right)^2+4\lambda_{HS}^2v^2v_R^2}\nonumber\\
&\simeq& 2\lambda v^2\left(1-\frac{\lambda_{HS}^2}{\lambda\,\lambda_S}\right),
\end{eqnarray}
where $v_R=\Lambda_R\simeq4$ TeV in our case.
The correlation between the values for $\lambda_S(\Lambda_R)$ and $\lambda_{HS}(\Lambda_R)$ for the correct Higgs mass is shown in FIG.~\ref{lmvalues},
which represents the predictions of $SU(2/2)$ at 4~TeV.
The plot shows some values ($0.15-0.25$) in the perturbative region. We can also find larger values for these couplings as long as $(1-\lambda_{HS}^2/\lambda\,\lambda_S)\geq0$, while $\lambda$ remains always small for the correct Higgs mass.

\section{ \normalsize{Fermions}}
The leptons can be incorporated in the $SU(2/1)$ or $SU(2/2)$ construction by taking advantage of the vector space isomorphism between the Clifford algebra and the exterior algebra. Defining the Dirac operator as $\slashed {D}=\slashed{\partial}\cdot \textbf{I}+\frac{g}{2}\slashed{\mathcal{J}}$,\footnote{Recall that we shifted $g$ to $g/2$ in the original construction to match the conventional SM notation. This is why we have $g/2$ in front of $\mathcal{J}$.} where $\slashed{\mathcal{J}}$ is $\mathcal{J}$ with the one-forms in it contracted \cite{Coquereaux:1990ev}, $\mathcal{L}_f=\overline{\psi} i\slashed D\psi$ gives the necessary terms (including the Yukawa terms) for both constructions.
Not surprisingly, we have a relation for the Yukawa couplings ($Y=g/\sqrt{2}$) from the embedding, just like the one we have for $\sin^2{\theta_W}$. This prediction of the Yukawa coupling universality is a common problem in the literature
and it should be lifted via some suitable mechanism. For example, there might exist some mixing with new degrees of freedom at around $4$ TeV which may change the running of the couplings and still satisfy the constraint at this scale.
%
%
%

\section*{\normalsize{Conclusion}}

In this Letter we have discussed an emergent superconnection formulation of the
EW sector of the SM and its minimal extension which accommodates
the observed Higgs mass based on the supergroups $SU(2/1)$ and $SU(2/2)$, respectively. The $SU(2/1)$ formalism unifies the Higgs and the gauge sectors (of the EW part) of the SM. It gives a geometric meaning to the low energy world, which also offers an explanation for the robustness of the SM. However, the model does not predict the Higgs mass correctly. Therefore, in this emergent geometric approach, we introduce \emph{new physics} in the form of $SU(2/2)$  which involves the left-right symmetric model ($SU(2)_L\times SU(2)_R\times U(1)_{B-L}$). Although this formalism does \emph{not} uniquely predict the Higgs mass (thus it is not the unique extension of $SU(2/1)$), we show that there is an available parameter space in this model which accommodates the observed mass of the Higgs.

This formalism predicts the scale of the onset of new physics (the left-right symmetric model) as 4 TeV. In addition to the usual implications of the TeV scale left-right symmetric model, it also predicts constraints, presumably valid at 4 TeV, which relate the quartic Higgs and Yukawa couplings to the gauge coupling of $SU(2)_L$. The latter brings the problem of Yukawa coupling universality which should be lifted via some suitable mechanism, e.g. taking into account mixing with heavy states.%

Given the observation made in Ref.~\cite{nondec} regarding the violation of decoupling
in the Higgs sector, and given the similarities between the superconnection approach and the Spectral SM,
this violation of decoupling in the Higgs sector
could be viewed in the context of non-commutative geometry
as indicating the mixing of the UV and IR degrees of freedom.
Similar UV/IR mixing is known in the simpler example of non-commutative field theory \cite{nc}
and is expected to appear in the more general context of non-perturbative quantum gravity~\cite{NPQG}.
In view of such non-decoupling, one could imagine
that the appearance of the Pati-Salam degrees of freedom (as well as the embedded SM
degrees of freedom) at low energy is essentially a direct manifestation of this UV/IR mixing.
Thus, on the one hand, the remnants of the UV physics could
be expected at a low energy scale of 4 TeV, and conversely, the Pati-Salam structure (and the embedded SM) might point to some unique features of the high energy physics of quantum gravity.
In this context, we should briefly mention the observations made in Ref.~\cite{dienes} about the special nature of the Pati-Salam group in certain constructions of
string vacua. This opens an exciting possibility of new experimental probes of fundamental short distance physics.


\section*{ \normalsize{Acknowledgements}}
We thank Lay Nam Chang, David Fairlie and Murat G\"{u}naydin for
interesting conversations.
UA, DM, TT are supported
by the U.S. Department of Energy
under contract DE-FG05-92ER40677, Task A.
TT is also supported by the World Premier International Research Center Initiative (WPI Initiative), MEXT, Japan.


\end{document}